\documentclass[12pt,twoside]{article}
\usepackage{correl}

\def\TheTitle{An introduction to many-body Green's functions in and out of equilibrium}
\def\TheHeading{Many-body Green's functions}
\def\TheAuthor{James K. Freericks}
\def\TheInstitute{Department of Physics, Georgetown University}
\def\TheAddress{Reiss Science Building, 37th and O Sts. NW, \\Washington, DC 20057, U.~S.~A.}
\def\TheChapter{~}           

\begin{document}
\MakeTitle           
\section{Introduction}

\index{Green's function}
This chapter is all about Green's functions. Many-body Green's functions are a marvelous tool to employ in quantum-mechanical calculations. They require a rather large set of mathematical machinery to develop their theory and a sophisticated computational methodology to determine them for anything but the simplest systems. But what do we really use them for?  Essentially, there are two main uses: one is to compute the thermodynamic expectation value of $c^\dagger c$, which allows us to determine all interesting single-particle expectation values, like the kinetic energy, the momentum distribution, etc. and two is to determine the many-body density of states, which tells us how the quantum states are distributed in momentum and energy. In large dimensions, they can also be used to determine some two-particle correlation functions like the optical conductivity, because the vertex corrections vanish. It seems like this is a lot of work to end up with only a few concrete results. But that is the way it is. No one has figured out any way to do this more efficiently. If you determine how to, fame and fortune are likely to follow!
\index{Green's function}

\index{Green's function!history}
Before jumping into the full glory, we will give just a brief history. This is one that focuses on my opinion of where critical ideas originated. It is not intended to be exhaustive or complete, and of course I may be wrong about where the different ideas came from. The equilibrium theory for Green's functions was developed primarily in the 1950s. The Lehmann representation~\cite{lehmann} was discovered in 1954. The Russian school developed much of the perturbative approach to Green's functions, which is summarized in the monograph of Abrikosov, Gorkov, and Dzyaloshinski~\cite{agd}. Joaquin Luttinger and John Ward developed a functional approach~\cite{luttinger_ward} and Matsubara determined how to solve for thermal Green's functions~\cite{matsubara}. But the reference I like the most for clarifying many-body Green's functions both in and out of equilibrium is the monograph by Baym and Kadanoff~\cite{baym_kadanoff}. Their approach is one we will follow, at least in spirit, in this chapter. Of course, Keldysh's perpsective~\cite{keldysh} was also important.  Serious numerical calculations of Green's functions (in real time and frequency) only began with the development of dynamical mean-field theory in the 1990s~\cite{dmft_review}. The nonequilibrium generalization began only in the mid 2000's~\cite{freericks1,freericks2,noneq_rmp_rev}, but it was heavily influenced by earlier work from Uwe Brandt~\cite{brandt_mielsch1,brandt_mielsch2,brandt_neq}.
\index{Green's function!history}

Now that we have set the stage for you as to where we are going, get ready for the ride. The theory is  beautiful, logical, abstract, and complex. Mastering it is a key in becoming a many-body theorist. We begin in section 2 with a discussion of equilibrium Green's functions focused on the Lehmann representation. Section 3, generalizes and unifies the formalism to the contour-ordered Green's function. Section 4 introduces the self-energy and the equation of motion. Section 5 illustrates how to include the external electric field. Section 6 introduces how one solves these problems within the nonequilibrium dynamical mean-field theory approach. Numerics are discussed in section 7, followed by examples in section 8. We conclude in section 9.

\section{Green's functions in equilibrium and the Lehmann representation}

\index{Green's function!equilibrium}
\index{Green's function!retarded}
\index{Green's function!advanced}
\index{Green's function!lesser}
\index{Green's function!greater}
We begin with the definition of four Green's functions in equilibrium: the retarded $G_{ij\sigma}^R(t)$, advanced $G_{ij\sigma}^A(t)$, lesser $G_{ij\sigma}^<(t)$ and greater $G_{ij\sigma}^>(t)$. These Green's functions are time-dependent complex thermal expectation values weighted by the thermal density matrix $\exp(-\beta{\mathcal H})/{\mathcal Z}$, with ${\mathcal Z}={\rm Tr}\exp(-\beta{\mathcal H})$ the partition function. They are defined in terms of the (anticommuting)  fermionic creation (annihilation) operators for an electron at site $i$ with spin $\sigma$: $c_{i\sigma}^\dagger$ ($c_{i\sigma}^{\phantom\dagger}$). Time dependence is handled in the Heisenberg picture, where the operators depend on time and the states do not, given by ${\mathcal O}(t)=U^\dagger(t){\mathcal O}U(t)$. For a time-independent Hamiltonian, the time-evolution operator $U(t)$ is simple and expressed as $U(t)=\exp(-i{\mathcal H}t)$, where we set $\hbar=1$. The four Green's functions then become
\begin{eqnarray}
G_{ij\sigma}^R(t)&=&-i\theta(t)\frac{1}{\mathcal Z}{\rm Tr}e^{-\beta{\mathcal H}}\{c_{i\sigma}^{\phantom\dagger}(t),c_{j\sigma}^\dagger(0)\}_+,\label{eq:GR_def}\\
G_{ij\sigma}^A(t)&=&i\theta(-t)\frac{1}{\mathcal Z}{\rm Tr}e^{-\beta{\mathcal H}}\{c_{i\sigma}^{\phantom\dagger}(t),c_{j\sigma}^\dagger(0)\}_+,\label{eq:GA_def}\\
G_{ij\sigma}^<(t)&=&i\frac{1}{\mathcal Z}{\rm Tr}e^{-\beta{\mathcal H}}c_{j\sigma}^\dagger(0)c_{i\sigma}^{\phantom\dagger}(t),\label{eq:G<_def}\\
G_{ij\sigma}^>(t)&=&-i\frac{1}{\mathcal Z}{\rm Tr}e^{-\beta{\mathcal H}}c_{i\sigma}^{\phantom\dagger}(t)c_{j\sigma}^\dagger(0).\label{eq:G>_def}
\end{eqnarray}
Here, $\theta(t)$ is the Heaviside unit step function, which vanishes for $t<0$, is equal to 1 for $t>0$ and we choose it to be $1/2$ for $t=0$, and the symbol $\{\cdot,\cdot\}_+$ denotes the anticommutator. Note that the fermionic operators satisfy the canonical anticommutation relations, namely $\{c_{i\sigma}^{\phantom\dagger},c_{j\sigma^\prime}^\dagger\}_+=\delta_{ij}\delta_{\sigma\sigma^\prime}$.
These Green's functions can also be defined in momentum space, with the obvious changes of the subscripts from real space to momentum space---the only significant modification is that the Green's functions will be {\it diagonal} in momentum for translationally invariant systems---which is all we will study in this chapter.
\index{Green's function!equilibrium}
\index{Green's function!retarded}
\index{Green's function!advanced}
\index{Green's function!lesser}
\index{Green's function!greater}

\index{Green's function!momentum-dependent}
\index{Green's function!retarded}
\index{Lehmann representation}
\index{density of states}
\index{Heisenberg representation}
These definitions may look like they come out of the blue, but the key point to realize is that when $t=0$, the lesser Green's function provides precisely the expectation value we need to be able to calculate any single-particle expectation value. We will see in a moment that the retarded Green's function also leads to the momentum-dependent spectral function and the density of states. The best way to understand and make sense of these Green's functions is with the Lehmann representation~\cite{lehmann}. This allows us to explicitly determine the Green's functions as functions of frequency via a Fourier transformation:
\begin{equation}
G_{ij\sigma}(\omega)=\int_{-\infty}^\infty dt e^{i\omega t}G_{ij\sigma}(t).\label{eq:FT}
\end{equation}
The trick is to insert the resolution of unity (${\mathbb I}=\sum_n|n\rangle\langle n|$ where $|n\rangle$ is an eigenstate of $\mathcal H$ such that ${\mathcal H}|n\rangle=E_n|n\rangle$) in between the creation and annihilation operators in the definitions of the Green's functions. We illustrate how this works explicitly for the retarded Green's function, where we have
\begin{equation}
G_{ij\sigma}^R(t)=-i\theta(t)\frac{1}{\mathcal Z}\sum_{mn} \left [ \langle m|e^{-\beta \mathcal H}c_{i\sigma}^{\phantom\dagger}(t)|n\rangle\langle n|c_{j\sigma}^\dagger |m\rangle+\langle m| e^{-\beta\mathcal H}c_{j\sigma}|n\rangle\langle n|c_{i\sigma}^{\phantom\dagger}(t)|n\rangle\right ].
\label{eq:lehmann1}
\end{equation}
We now use the fact that these states are eigenstates of $\mathcal H$ and the Heisenberg representation for the time dependent operators to find
\begin{equation}
G_{ij\sigma}^R(t)=-i\theta(t)\sum_{mn} \frac{e^{-\beta E_m}}{\mathcal Z}\left [ e^{i(E_m-E_n)t}
\langle m| c_{i\sigma}^{\phantom\dagger}|n\rangle\langle n|c_{j\sigma}^\dagger|m\rangle+
e^{-i(E_m-E_n)t}\langle m|c_{j\sigma}^\dagger|n\rangle\langle n|c_{i\sigma}^{\phantom\dagger}|m\rangle\right ].
\label{eq:lehmann2}
\end{equation}
Next, we interchange $m\leftrightarrow n$ in the second term to give
\begin{equation}
G_{ij\sigma}^R(t)=-i\theta(t)\frac{1}{\mathcal Z}\sum_{mn} \left ( e^{-\beta E_m}+e^{-\beta E_n}\right )
e^{i(E_m-E_n)t}\langle m |c_{i\sigma}^{\phantom\dagger}|n\rangle\langle n|c_{j\sigma}^\dagger|m\rangle.
\label{eq:lehmann3}
\end{equation}
It is now time to do the Fourier transform. We achieve this by shifting $\omega\to\omega+i0^+$ to make the integrand vanish at the large time limit, which finally results in
\begin{equation}
G_{ij\sigma}^R(\omega)=\frac{1}{\mathcal Z}\sum_{mn}\frac{e^{-\beta E_m}+e^{-\beta E_n}}{\omega+E_m-E_n+i0^+}\langle m |c_{i\sigma}^{\phantom\dagger}|n\rangle\langle n|c_{j\sigma}^\dagger|m\rangle.
\label{eq:lehmann4}
\end{equation}
This is already of the form that we can discover some interesting things. If we recall the Dirac relation $\frac{1}{\omega+i0^+}=\frac{\mathcal P}{\omega}-i\pi\delta(\omega)$ (with $\mathcal P$ denoting the principal value), then we find the local density of states via
\begin{equation}
A_{i\sigma}(\omega)=-\frac{1}{\pi}{\rm Im}G_{ii\sigma}(\omega)=\sum_{mn}\delta(\omega+E_m-E_n)\frac{ e^{-\beta E_m}+e^{-\beta E_n}}{\mathcal Z}|\langle m|c_{i\sigma}|n\rangle|^2.
\label{eq:dos}
\end{equation}
This density of states satisfies a sum rule. In particular,
\begin{equation}
\int_{-\infty}^\infty d\omega A_{i\sigma}(\omega)=\frac{1}{\mathcal Z}\sum_n e^{-\beta E_n}\left ( c_{i\sigma}^{\phantom\dagger}c_{i\sigma}^\dagger+c_{i\sigma}^\dagger c_{i\sigma}^{\phantom\dagger}\right )=1,
\label{eq:sum_rule}
\end{equation}
which follows from the anticommutation relation of the fermionic operators; note that you can also see this directly by taking the limit $t\to 0^+$ in Eq.~(\ref{eq:GR_def}). This result holds in momentum space too:
the retarded Green's function becomes
\begin{eqnarray}
G_{k\sigma}^R(t)&=&-i\theta(t)\frac{1}{\mathcal Z}{\rm Tr}e^{-\beta \mathcal H}\left \{ c_{k\sigma}^{\phantom\dagger}(t),c_{k\sigma}^\dagger(t)\right \}_+\\
G_{k\sigma}^R(\omega)&=&\frac{1}{\mathcal Z}\sum_{mn}\frac{e^{-\beta E_m}+e^{-\beta E_n}}{\omega+E_m-E_n+i0^+}\langle m|c_{k\sigma}^{\phantom\dagger}|n\rangle\langle n|c_{k\sigma}^\dagger|m\rangle,
\label{eq: GR_mom}
\end{eqnarray}
with the operators being the momentum space operators, determined by a Fourier transform from real space:
\begin{equation}
c_{k\sigma}^{\phantom\dagger}=\frac{1}{|\Lambda|}\sum_{j\in\Lambda}c_{j\sigma}^{\phantom\dagger}e^{-i{\bf k}\cdot{\bf R}_j}~~~~{\rm and}~~~~
c_{k\sigma}^{\dagger}=\frac{1}{|\Lambda|}\sum_{j\in\Lambda}c_{j\sigma}^{\phantom\dagger}e^{i{\bf k}\cdot{\bf R}_j}.
\end{equation}
Here, the lattice with periodic boundary conditions is denoted by $\Lambda=\{{\bf R}_i\}$ and $|\Lambda|$ is the number of lattice sites in the system. We usually do not use a bold font for the momentum label in the subscript of the creation or annihilation operator; one should be able to figure out from the context whether we are working in real space or momentum space.
The momentum-space Green's functions have a similarly defined spectral function $A_{k\sigma}(k,\omega)=-{\rm Im}G^R_{k\sigma}(\omega)/\pi$, which also satisfies a sum rule given by the integral over all frequency being equal to one. One can easily show that the advanced Green's function in frequency space is equal to the Hermitian conjugate of the retarded Green's function, because we need to shift $\omega\to\omega-i0^+$ to control the integrand at the lower limit. Hence, we have $G^A_{ij\sigma}(\omega)=G^{R*}_{ji\sigma}(\omega)$.
\index{Green's function!momentum-dependent}
\index{Green's function!retarded}
\index{Lehmann representation}
\index{density of states}
\index{Heisenberg representation}

\index{Green's function!momentum-dependent}
\index{Green's function!lesser}
\index{Green's function!greater}
\index{Lehmann representation}
\index{spectral function}
\index{fluctuation-dissipation theorem}
We also sketch how the calculation works for the lesser and greater Green's functions. Here, we have to control the integrand at both endpoints of the integral. To do so, we split it at 0  and introduce the appropriate infinitesimal convergence factor in each piece. The rest of the calculation proceeds as above and one finds that the two pieces have their imaginary parts add, so we finally obtain:
\begin{equation}
G_{ij\sigma}^<(\omega)=2i\pi\frac{1}{\mathcal Z}\sum_{mn}\delta(\omega+E_m-E_n)e^{-\beta E_n}\langle m|c_{i\sigma}^{\phantom\dagger}|n\rangle\langle n|c^\dagger_{j\sigma}|m\rangle.
\end{equation}
The greater Green's function similarly becomes
\begin{equation}
G_{ij\sigma}^>(\omega)=-2i\pi\frac{1}{\mathcal Z}\sum_{mn}\delta(\omega+E_m-E_n)e^{-\beta E_m}\langle m|c_{i\sigma}^{\phantom\dagger}|n\rangle\langle n|c^\dagger_{j\sigma}|m\rangle.
\end{equation}
Note that we can use the delta function to immediately infer an important identity, namely that 
$G^<_{ij\sigma}(\omega)=-\exp(-\beta\omega)G^>_{ij\sigma}(\omega)$. Combining this result with the fact that ${\rm Im}G_{ij\sigma}^R(\omega)={\rm Im}[G_{ij\sigma}^<(\omega)-G_{ij\sigma}^>(\omega)]/2$, we finally learn that
\begin{equation}
G_{ij\sigma}^<(\omega)=-2 if(\omega){\rm Im}G_{ij\sigma}^R(\omega)~~~{\rm and}~~~
G_{ij\sigma}^>(\omega)=2 i[1-f(\omega)]{\rm Im}G_{ij\sigma}^R(\omega).
\end{equation}
This equilibrium relationship is sometimes called the fluctuation-dissipation theorem.
\index{Green's function!momentum-dependent}
\index{Green's function!lesser}
\index{Green's function!greater}
\index{Lehmann representation}
\index{spectral function}
\index{fluctuation-dissipation theorem}

\section{Green's functions out of equilibrium and the ``contour''}

\vskip -1 in
\begin{figure}[htb!]
 \centering
\begin{minipage}[e]{0.49\textwidth}
\includegraphics[width=1.2\textwidth]{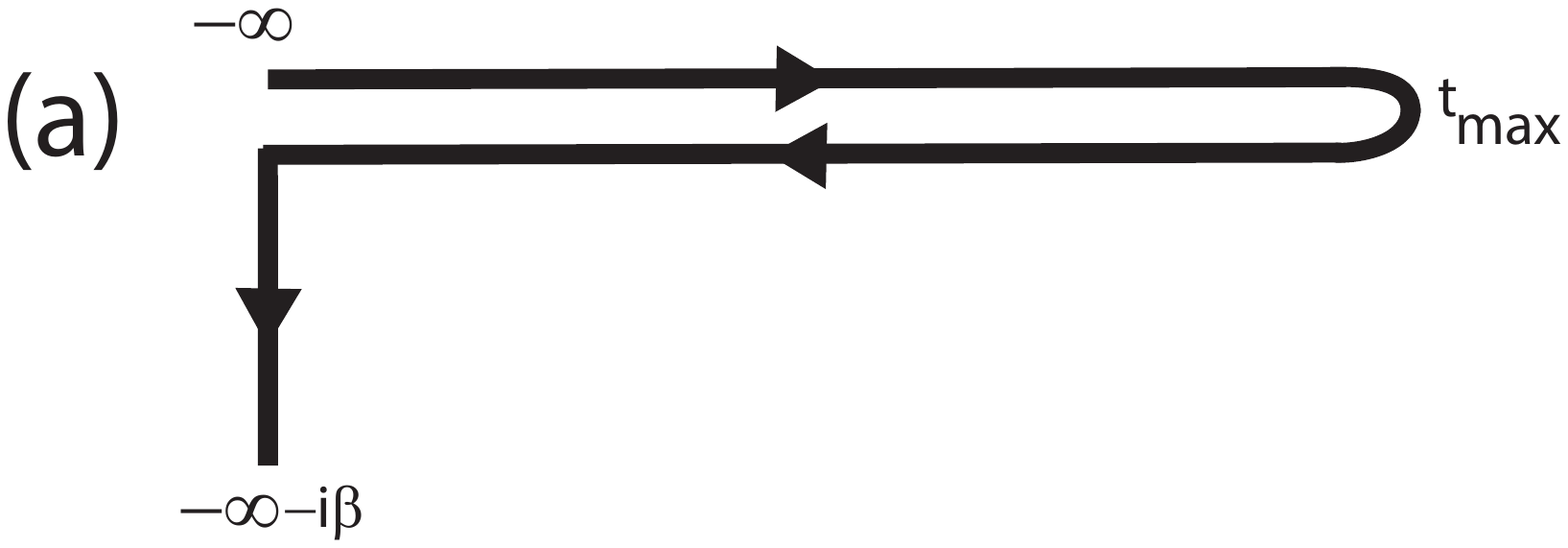}
\end{minipage}
\begin{minipage}[e]{0.49\textwidth}
\vskip -1.8in
\includegraphics[width=1.2\textwidth]{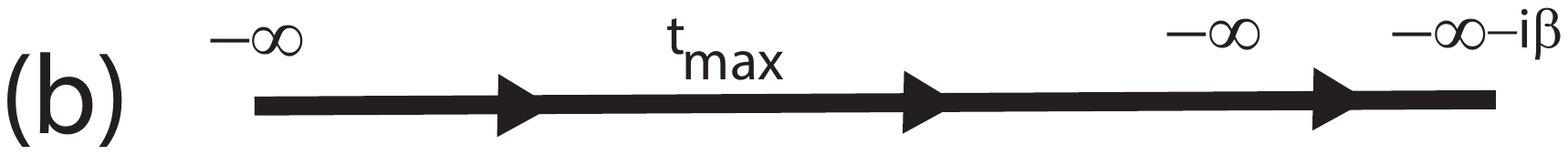}
\end{minipage}
\vskip -2.2in
 \caption{(a) Kadanoff-Baym-Keldysh contour used in the contruction of the contour-ordered Green's function. The contour starts at $-\infty$, runs out to the maximum of $t$ and $t^\prime$, runs back to $-\infty$ and then runs parallel to the imaginary axis for a distance equal to $\beta$. (b) The contour can be ``stretched'' into a straight line as indicated here, which is convenient for properly implementing time-ordering along the contour.}
 \label{fig:contour}
\end{figure}

\index{Kadanoff-Baym-Keldysh contour}
\index{Green's function!time representation}
Now we move onto nonequilibrium where the Hamiltonian is time-dependent ${\mathcal H}(t)$. The evolution operator depends on two times and becomes $U(t,t^\prime)={\mathcal T}_t \exp\left [-i\int_{t^\prime}^t d\bar t {\mathcal H}(\bar t)\right ]$. It satisfies the semigroup property $U(t,t^{\prime\prime})U(t^{\prime\prime},t^\prime)=U(t,t^\prime)$, with $U(t,t)={\mathbb I}$ and $i\partial_t U(t,t^\prime)={\mathcal H}(t)U(t,t^\prime)$. The Green's functions in nonequilibrium are defined by the same equations as before, namely Eqs.~(\ref{eq:GR_def}--\ref{eq:G>_def}), except now they depend on two times, with $t^\prime$ being the argument of the $c^\dagger$ operator. All of these Green's functions require us to evaluate one of two matrix elements. We examine one of them here and assume $t>t^\prime$ for concreteness:
\begin{eqnarray}
\frac{1}{\mathcal Z}{\rm Tr}e^{-\beta \mathcal H}c_{i\sigma}^{\phantom\dagger}(t) c_{j\sigma}^\dagger(t^\prime)&\hskip -0.1in =&\hskip -0.1in
\frac{1}{\mathcal Z}{\rm Tr}e^{-\beta \mathcal H}U^\dagger(t,-\infty)c_{i\sigma}^{\phantom\dagger}U(t,-\infty)U^\dagger(t^\prime,-\infty)c_{j\sigma}^\dagger U(t^\prime,-\infty)\\
&\hskip -0.1in =&\hskip -0.1in \frac{1}{\mathcal Z}{\rm Tr}e^{-\beta \mathcal H}U(-\infty,t)c_{i\sigma}^{\phantom\dagger}U(t,t^\prime)c_{j\sigma}^\dagger U(t^\prime,-\infty)
\end{eqnarray}
where we used the facts that $U^\dagger(t_1,t_2)=U(t_2,t_1)$ and the semigroup property to show that $U(t,-\infty)U^\dagger(t^\prime,-\infty)=U(t,t^\prime)$. The time-evolution operators, including time evolution in the imaginary axis direction for the $\exp(-\beta{\mathcal H})$ term, can be seen to all live on the so-called Kadanoff-Baym-Keldysh contour, which is shown in Fig.~\ref{fig:contour}. Starting with time at $-\infty$, we evolve forward to $t^\prime$, then apply the $c^\dagger$, evolve forward to time $t$, apply the $c$ operator, and then evolve {\it backwards} in time to $-\infty$. The lesser and greater Green's functions are easily determined in this fashion. The retarded or advanced Green's functions can then be found by taking the appropriate differences of greater and lesser functions        with the convention that $t>t^\prime$ for the retarded case and $t<t^\prime$ for the advanced case. Now that we have these Green's functions, we can generalize the definition of the Green's function to allow both times to lie anywhere on the contour.  We take the time-ordered Green's function, with the time-ordering taking place along the contour, and we call it the contour-ordered Green's function:
\begin{equation}
G^c_{ij\sigma}(t,t^\prime)=\left \{ \begin{matrix}
-i\frac{1}{\mathcal Z}{\rm Tr} e^{-\beta\mathcal H}\,c_{i\sigma}^{\phantom\dagger}(t)c_{j\sigma}^\dagger(t^\prime)~~{\rm for}~t>_ct^\prime\\
~\\
i\frac{1}{\mathcal Z}{\rm Tr} e^{-\beta\mathcal H}\,c_{j\sigma}^\dagger(t^\prime)c_{i\sigma}^{\phantom\dagger}(t)~~~~~{\rm for}~t<_ct^\prime.
\end{matrix}\right .,
\label{eq:gc_def}
\end{equation}
where the $c$ subscript on the less than or greater than symbol is to denote whether one time is ahead of or behind the other on the contour, regardless of the numerical values of $t$ and $t^\prime$.
This Green's function is the workhorse of nonequilibrium many-body physics---we use it to calculate anything that can be calculated with the single-particle Green's functions.
\index{Kadanoff-Baym-Keldysh contour}
\index{Green's function!time representation}

Note that in these lectures, we will work with the contour-ordered Green's function itself, which is a continuous matrix operator in time. In many other works, the Green's function is further decomposed into components where the times lie on the same or different branches of the real contour ($2\times 2$ matrix)~\cite{keldysh} or of the two real and one imaginary branches of the contour ($3\times 3$ matrix)~\cite{wagner}. In the $3\times 3$ case, one has real, imaginary and mixed Green's functions, the last ones determining the so-called initial correlations. In the $2\times 2$ case, it is common to transform the matrix to make it upper triangular, expressing the retarded, advanced and so-called Keldysh Green's functions. In general, I find that these decompositions and transformations make the material more difficult than it  actually is and it is best to defer using them until one has gained some mastery of the basics. Here, we will work solely with the contour-ordered functions.

\section{The self-energy and the equation of motion}

\index{Green's function!contour-ordered}
\index{equation of motion}
\index{self-energy}
The next step in our analysis is to derive the equation of motion. We start from the definition of the contour-ordered Green's function in Eq.~(\ref{eq:gc_def}) and we employ a unit step function on the contour $\theta_c(t,t^\prime)$ which is one for $t$ ahead of $t^\prime$ on the contour and zero for $t$ behind $t^\prime$. This theta function can be used to impose the two conditions. Its derivative with respect to time becomes a delta function on the contour $\partial_t\theta_c(t,t^\prime)=\delta_c(t,t^\prime)$. The delta function on the contour satisfies 
\begin{equation}
\int_c dt \delta(t,t^\prime)f(t^\prime)=f(t),
\end{equation}
for any time $t$ on the contour (including those on the imaginary spur). Now a simple derivative of the Green's function with respect to time yields the equation of motion (we find it convenient to illustrate this in momentum space)
\begin{eqnarray}
&~&i\frac{\partial}{\partial t} G_{k\sigma}^c(t,t^\prime)=\frac{1}{\mathcal Z}{\rm Tr}e^{-\beta \mathcal H}\left \{c_{k\sigma}^{\phantom\dagger}(t),c_{k\sigma}^\dagger(t)\right\}\delta_c(t,t^\prime)\\
&+&i\theta_c(t,t^\prime)\frac{1}{\mathcal Z}{\rm Tr}e^{-\beta\mathcal H}\left [{\mathcal H}(t),c_{k\sigma}^{\phantom\dagger}(t)\right ]c_{k\sigma}^\dagger(t^\prime)-i\theta_c(t^\prime,t)\frac{1}{\mathcal Z}{\rm Tr}e^{-\beta\mathcal H}c_{k\sigma}^\dagger(t^\prime)\left [{\mathcal H}(t),c_{k\sigma}^{\phantom\dagger}(t)\right ]\nonumber\\
&=&\delta_c(t,t^\prime)+[\epsilon_k(t)-\mu]G_{k\sigma}^c(t,t^\prime)
+\int_cdt^{\prime\prime}\Sigma_c(t,t^{\prime\prime})G_{k\sigma}^c(t^{\prime\prime},t^\prime).
\label{eq:eom}
\end{eqnarray}
Here, we assumed that the Hamiltonian (in the Schr\"odinger picture) takes the form
\begin{equation}
{\mathcal H_s}(t)=\sum_{k\sigma}[\epsilon_k(t)-\mu]c_{k\sigma}^\dagger c_{k\sigma}^{\phantom\dagger}+V(t)
\end{equation}
where $\epsilon_k(t)$ is a time-dependent bandstructure and $\mu$ is the chemical potential. The symbol $V(t)$ is the potential energy operator of the Hamiltonian. The commutator of the potential with the fermionic destruction operator is complicated---we introduce the contour-ordered self-energy $\Sigma_c$ to describe that cumbersome term---which can be viewed as the definition of the self-energy. The self-energy appears as a convolution with the Green's function (over the entire contour) in Eq.~(\ref{eq:eom}). The equation of motion can be rearranged to look like
\begin{equation}
\int_c dt^{\prime\prime}\left \{ \left [ i\frac{\partial}{\partial t}-\epsilon_k(t)+\mu\right ]\delta_c(t,t^{\prime\prime})+\Sigma_c(t,t^{\prime\prime})\right \}G_{k\sigma}^c(t^{\prime\prime},t^\prime)=\delta_c(t,t^\prime).
\end{equation}
As we mentioned before, the Green's functions are continuous matrix operators. Seen in that light, it should be clear that the object inside the curly brackets is the matrix inverse of the Green's function (since matrix multiplication, implied by the integration over the contour, has the matrix defined by the curly brackets multiplied by $G^c$ and equaling the identity matrix).
\index{Green's function!contour-ordered}
\index{equation of motion}
\index{self-energy}

At this stage it may look like we are pulling the wool over your eyes. In some sense we are. We have hidden the difficult question of how to evaluate the commutator of the potential with the destruction operator, by simply introducing a self-energy which makes the complications go away. This really would lead us nowhere if we did not have another means by which we can determine the self-energy. Fortunately we do, and it will be introduced below. You will have to wait to see how that fits in. All of the challenge in many-body physics boils down to this question of ``how can we calculate the self-energy?''

Now is also a good time to discuss the issue of how do we work with these complicated objects? If you look carefully, you will see we are working with matrices. Admittedly they are continuous matrix operator equations, but like any continuous object, we discretize them when we want to work with them on a computer. And indeed, that is what we will do here when we describe the DMFT algorithm below. We always extrapolate the discretization to zero to recover the continuous matrix operator results. It is important to recognize these facts now, as it will make your comfort level with the approach much higher as we develop the subject further.

\section{Incorporating an external field and the issue of gauge invariance}

I always found it odd in dealing with fields in quantum mechanics to learn that we do not always respect Maxwell's equations. This should not come as a surprise to you, because Maxwell's equations are relativistically invariant, and quantum mechanics only becomes relativistically invariant with a properly constructed relativistic quantum field theory. Nevertheless, we choose to neglect even more aspects of Maxwell's equations than just relativistic invariance. This becomes necessary because the quantum problem cannot be solved unless we do this (and, fortunately, the neglected terms are small).

The main issue we face is that we want the field to be spatially uniform but time varying. Maxwell says such fields do not exist, but we press on anyway. You see, a spatially uniform field means that I still have translation invariance and that is needed to make the analysis simple enough that it can be properly carried out. We justify this in part because optical wavelengths are much larger than lattice spacings and because magnetic field effects are small due the extra $1/c$ factors multiplying them.
Even with this simplification, properly understanding how to introduce a large amplitude electric field is no  easy task. 

\index{Peierls' substitution}
The field will be described via the Peierls' substitution~\cite{peierls}. This is accomplished by shifting $k\to k-eA(t)$, which becomes $k+eEt$, for a constant (dc) field. This seems like the right thing to do, because it parallels the minimal substitution performed in single-particle quantum mechanics when incorporating an electric field, but we have to remember that we are now working with a system projected to the lowest lying $s$-orbital band, so it may not be obvious that this remains the correct thing to do. Unfortunately, there is no rigorous proof that this result is correct, but one can show that if we work in a single band (neglecting interband transitions) and have inversion symmetry in the lattice, then the field generated by a scalar potential, and the field generated by the equivalent vector potential will yield identical noninteracting Green's functions~\cite{joura}. We then assume that adding interactions to the system does not change this correspondence, so we have an approach that is exact for a single-band model. One should bear in mind that there may be subtle issues with multiple bands and the question of how to precisely work with gauge-invariant objects in mutliband models has not been completely resolved yet. Fortunately, for model system calculations, working with a single band is often sufficient to illustrate the nonquilibrium phenomena. This is what we do here.
\index{Peierls' substitution}

\index{Falicov-Kimball model}
\index{Peierls' substitution}
It is now time to get concrete in describing the Hamiltonian we will be working with, which is the spinless Falicov-Kimball model~\cite{falicov_kimball}. This model is nice to work with because it does possess strongly correlated physics (Mott transition and order-disorder transitions), but it can be solved exactly within DMFT in both equilibrium and nonequilibrium. Hence, it provides a nice test case to illustrate the approach we take here. The Hamiltonian, in an electric field is
\begin{equation}
{\mathcal H}_S(t)=\sum_k\left [ \epsilon(k-eA(t))-\mu\right ]c_k^\dagger c_k^{\phantom\dagger}
+E_f\sum_i f_i^\dagger f_i^{\phantom\dagger}+U\sum_i c_i^\dagger c_i^{\phantom\dagger} f_i^\dagger f_i^{\phantom\dagger},
\label{eq:HFK}
\end{equation}
where we employed the Peierls' substitution in the bandstructure. Here, the conduction electrons ($c$) are spinless and do not interact with themselves. They interact instead with the localized electrons ($f$), when both electrons are situated on the same lattice site---$E_f$ is the local site energy of the $f$-electrons. Note that we use $k$ for momentum and $i$ for position in ${\mathcal H}$. All of the time dependence is in the kinetic-energy term, because the field does not couple to the localized electrons (due to the fact that they do not hop). Hence, we have an effective single-band model and the discussion above about the applicability of the Peierls' substitution applies.
\index{Falicov-Kimball model}
\index{Peierls' substitution}

\index{Green's function!nonequilibrium!noninteracting}
It is instructive at this stage to determine the noninteracting Green's function exactly. We do this by determining the Heisenberg creation and annihilation operators from which we form the Green's function. In the case where $U=0$, one can immediately find that that the destruction operator (in the Heisenberg picture) becomes
\begin{equation}
c_{k\sigma}^{\phantom\dagger}(t)=e^{i\int_{-\infty}^t d\bar t [\epsilon(k-eA(\bar t)-\mu]c_{k\sigma}^\dagger c_{k\sigma}^{\phantom\dagger}}
c_{k\sigma}e^{-i\int_{-\infty}^t d\bar t [\epsilon(k-eA(\bar t)-\mu]c_{k\sigma}^\dagger c_{k\sigma}^{\phantom\dagger}}
=e^{-i\int_{-\infty}^t d\bar t [\epsilon(k-eA(\bar t))-\mu]}c_{k\sigma}^{\phantom\dagger}.
\label{eq:destruct_t}
\end{equation}
There is no need for time ordering here, because the Hamiltonian commutes with itself at different times. {\it This simplification is what allows one to solve the problem easily.}  By taking the Hermitian conjugate of Eq.~(\ref{eq:destruct_t}), one can substitute in the definitions of the Green's functions and immediately find that
\begin{equation}
G_{k\sigma}^<(t,t^\prime)=i\frac{1}{\mathcal Z}{\rm Tr}e^{-\beta\mathcal H}e^{-i\int_{t^\prime}^t d\bar t \left [ \epsilon(k-eA(\bar t))-\mu\right ]}c_{k\sigma}^\dagger c_{k\sigma}^{\phantom\dagger}=ie^{-i\int_{t^\prime}^t d\bar t \left [ \epsilon(k-eA(\bar t))-\mu\right ]}\langle n_{k\sigma}\rangle,
\end{equation}
where $\langle n_{k\sigma}\rangle={\rm Tr}\exp[-\beta{\mathcal H}]c_{k\sigma}^\dagger c_{k\sigma}^{\phantom\dagger}$ is the {\it initial equilibrium} momentum distribution. We similarly find that the noninteracting retarded Green's function in nonequilibrium is
\begin{equation}
G_{k\sigma}^R(t.t^\prime)=-i\theta(t-t^\prime)e^{-i\int_{t^\prime}^t d\bar t \left [ \epsilon(k-eA(\bar t))-\mu\right ]};
\end{equation}
there is no equilibrium expectation value here because the anticommutator of the equal-time fermionic operators is 1. To find the corresponding local Green's functions, we simply sum these expressions over all momenta.
\index{Green's function!nonequilibrium!noninteracting}

\index{gauge invariance}
\index{Green's function!gauge-invariant}
We next discuss the issue of gauge invariance. This issue is important, because we are required to perform calculations in a specific gauge but any physically measurable quantity must be gauge-invariant. So how does this come about? To begin, one should notice that the Peierls' substitution guarantees that any quantity that is local, like the current, is gauge invariant, because the sum over all momenta includes every term, even if the momenta are shifted due to the Peierls substitution. Momentum-dependent quantities, however, appear to depend directly on the vector potential. To properly formulate a gauge-invariant theory requires us to go back and formulate the problem (using a complete basis set, not a single-band downfolding) properly to illustrate the gauge-invariant nature of the observables. No one has yet been able to do this for a system with a nontrivial bandstructure. For a single-band model, we instead directly enforce gauge invariance using the so-called gauge-invariant Green's functions, introduced by Bertoncini and Jauho~\cite{bertoncini_jauho}. These Green's functions are constructed to be manifestly gauge invariant---the change in the phase of the creation and destruction operators is chosen to precisely cancel the change in the phase from the gauge transformation. We determine them in momentum space by making a shift
\begin{equation}
k\to k(t,t^\prime)=k+\int_{-\frac{1}{2}}^{\frac{1}{2}}d\lambda\, A\left ( \frac{t+t^\prime}{2}+\lambda(t-t^\prime)\right ).
\end{equation}
Note that one needs to be careful in evaluating this integral in the limit as $t\to t^\prime$. In this limit, we find $k\to k+A(t)$.
The explicit formula for the gauge-invariant Green's function, which uses a tilde to denote that it is gauge-invariant,  is
\begin{equation}
\tilde G_{k\sigma}(t,t^\prime)=G_{k(t,t^\prime)\sigma}(t,t^\prime).
\end{equation}
The demonstration of the gauge-invariance is given in Ref.~\cite{bertoncini_jauho}. Note that the local Greens functions are always manifestly gauge-invariant because the map $k\to k(t,t^\prime)$ is a ``one-to-one and onto'' map for any fixed set of times.
\index{gauge invariance}
\index{Green's function!gauge-invariant}

\index{current operator}
We end this section with a discussion of the current operator and how it is related to gauge invariance.
The current operator is calculated in a straightforward way via the commutator of the Hamiltonian with the charge polarization operator. We do not provide the details here, but simply note that they can be found in most many-body physics textbooks. The end result is that
\begin{equation}
j=\sum_\sigma\sum_k v_kc_{k\sigma}^\dagger c_{k\sigma}^{\phantom\dagger}.
\end{equation}
Here, we have the band velocity is given by $v_k=\nabla \epsilon_k$. The gauge-invariant form of the expectation value for the current is then
\begin{equation}
\langle j\rangle =-i\sum_\sigma\sum_k v_k \tilde G_{k\sigma}^<(t,t).
\end{equation}
We can re-express in terms of the original Green's functions, by shifting $k\to k-A(t)$. This yields the equivalent formula
\begin{equation}
\langle j\rangle = -i\sum_\sigma\sum_k v_{k-A(t)}G_{k\sigma}^<(t,t).
\end{equation}
As you can easily see, these two results are identical by simply performing a uniform shift of the momentum in the summation. Generically, this is how different expectation values behave---in the gauge-invariant form, one evaluates them the same as in linear response, but uses the gauge-invariant Green's functions, while in the non-gauge-invariant form, one evaluates in a form similar to linear response, but appropriately shifts the momentum in the terms that multiply the Green's functions (which is a beyond-linear-response ``correction'').
\index{current operator}

\section{Nonequilibrium dynamical mean-field theory}

\index{dynamical mean-field theory!nonequilibrium}
The premise of dynamical mean-field theory (DMFT) is that the self-energy of an impurity in an appropriately chosen time-dependent field is the same as the self-energy on the lattice. In general, we know this cannot be true, because the lattice self-energy depends on momentum, and hence is not local. But, if we choose the hopping matrix elements to scale like $t=t^*/2\sqrt{d}$ for a $d$-dimensional lattice, then one can rigorously show that in the limit $d\to\infty$, the self-energy does indeed become local~\cite{metzner}. So, in this limit, the DMFT assumption holds and it provides another limiting case where one can exactly solve the many-body problem (the other limit being that of $d=1$).
\index{dynamical mean-field theory!nonequilibrium}

\index{dynamical mean-field theory!nonequilibrium}
\index{dynamical mean-field theory!iterative algorithm}
The physical picture to keep in mind, for DMFT, is to think of looking at what happens just on a single site  of the lattice. As a function of time, we will see electrons hop onto and off of the site. If we can adjust the time-dependent field for the impurity in such a way that it produces the same behavior in time on the impurity, then the impurity site will look like the lattice. This motivates the following iterative algorithm~\cite{jarrell} to solve these models: (1) Begin with a guess for the self-energy ($\Sigma=0$ often works); (2) sum the momentum-dependent Green's function over all momentum to determine the local Green's function
\begin{equation}
G^c_{ii\sigma}(t,t')=\sum_k \left \{\left [ G_{k\sigma}^{c}(U{=}0)\right ]^{-1}-\Sigma\right \}^{-1}_{t,t^\prime},
\end{equation}
with $G_{k\sigma}^{c}(U{=}0)$ the noninteracting momentum-dependent contour-ordered Green's function on the lattice and the subscript denoting the $t,t^\prime$ element of the matrix inverse) ; (3) Use Dyson's equation to determine the effective medium $G_0$ via
\begin{equation}
\left [G_{0\sigma}^c\right ]^{-1}_{t,t^\prime}=\left [G_{ii\sigma}^c\right ]^{-1}_{t,t^\prime}+\Sigma^c(t,t^\prime);
\label{eq:dyson_eff_med}
\end{equation}
(4) solve the impurity problem in the extracted effective medium (which determines the time evolution on the impurity); (5) extract the self-energy for the impurity using an appropriately modified Eq.~(\ref{eq:dyson_eff_med}); and (8) repeat steps (2--7) until the self-energy has converged to the desired level of accuracy. The difference from the conventional iterative DMFT algorithm is that the objects worked with are now continuous matrix operators in time rather than simple functions of frequency. Sometimes this creates a roadblock for students to follow how the algorithm works, but there is no reason why it should. Finally, we note that because each inverse of a Green's function here implies a matrix inversion, we will want to organize the calculation in such a way as to minimize the inversions. This will be explained in more detail when we discuss numerical implementations below.
\index{dynamical mean-field theory!nonequilibrium}
\index{dynamical mean-field theory!iterative algorithm}

\index{bandstructure!hypercubic lattice}
One of the simplifications in equilibrium  DMFT is that the fact that the self-energy has no momentum dependence allows us to perform the summation over momentum via a single integral over the noninteracting density of states. Things are not quite as simple for the nonequilibrium case, as we now discuss. The bandstructure for a hypercubic lattice in $d\to\infty$ is given by
\begin{equation}
\epsilon_k=-\lim_{d\to\infty}\frac{t^*}{\sqrt{d}}\sum_{i=1}^d\cos(k_i).
\end{equation}
The central limit theorem tells us that $\epsilon_k$ is distributed according to a Gaussian distribution via
\begin{equation}
\rho(\epsilon)= \frac{1}{\sqrt{\pi}t^*}e^{-\epsilon^2}.
\end{equation}
In nonequilibrium, we have a second band energy to work with 
\begin{equation}
\bar\epsilon_k=-\lim_{d\to\infty}\frac{t^*}{\sqrt{d}}\sum_{i=1}^d\sin(k_i),
\end{equation}
because the Peierls substitution shifts $\cos(k_i)\to\cos(k_i)\cos(A_i(t))+\sin(k_i)\sin(A_i(t))$. The joint density of states becomes the product of two Gaussians, given by
\begin{equation}
	\rho(\epsilon,\bar\epsilon)=\frac{1}{\pi t^{*2}}e^{-\frac{\epsilon^2}{t^{*2}}-\frac{\bar\epsilon^2}{t^{*2}}}.
\end{equation}
This second band energy can be thought of as the projection of the band velocity onto the direction of the electric field. Hence, the computation of the local Green's function in step (2) is complicated by requiring a two-dimensional integration rather than a one-dimensional integration. This does make the numerics require significant additional resources, especially because the integrands are matrix-valued. 
\index{bandstructure!hypercubic lattice}

\index{dynamical mean-field theory!impurity solver}
The most challenging part of the algorithm is step (4)---solving the impurity problem in nonequilibrium. Unfortunately, there are few techniques available to do this. Monte Carlo methods suffer from the so-called phase problem, which might be able to be tamed using expansions about a perturbative solution and restricting updates to be nearly local in time via the inchworm algorithm~\cite{inchworm}. The other choice is perturbation theory either in the interaction (which often does not work so well, except for electron-phonon coupling) or the hybridization of the impurity model (which works well at half filling). Here, we make a different choice, and choose a simplified model that can be solved exactly, the so-called spinless Falicov-Kimball model~\cite{falicov_kimball}:
\begin{equation}
{\mathcal H}=\sum_{ij}t_{ij}c_i^\dagger c_j^{\phantom\dagger}-\mu\sum_i c_i^\dagger c_i^{\phantom\dagger}+E_f\sum_i f_i^\dagger f_i^{\phantom\dagger}+U\sum_i c_i^\dagger c_i^{\phantom\dagger} f_i^\dagger f_i^{\phantom\dagger}.
\label{eq:ham_fk}
\end{equation}
This model involves the interaction of conduction electrons ($c$) and localized electrons ($f$). The conduction electrons can hop on the lattice (we usually take the hopping only between nearest neighbors on a hypercubic lattice in the $d\to\infty$ limit) and have a chemical potential $\mu$. The localized electrons have a site energy $E_f$. Both electrons interact with an on-site interaction $U$.
\index{dynamical mean-field theory!impurity solver}

\index{Falicov-Kimball model}
The Falicov-Kimball model describes a rich set of physics, but it should not be viewed as a paradigm for all strongly correlated electrons. This is because it has some aspects that are not seen in more common models like the Hubbard model. For example, it is never a Fermi liquid when $U\ne 0$, the conduction electron density of states is independent of temperature in the normal state, and it is never a superconductor. But, it does display a lot of rich physics including a Mott metal-insulator transition, ordered charge-density-wave phases at low temperature and even phase separation when the densities of the two species are far enough apart. In addition, the $f$-electron density of states does exhibit strong temperature dependence, as expected in strongly correlated systems. It also displays orthogonality catastrophe behavior. The main interest in the nonequilibrium context has focused on the fact that it has a Mott transition. The most important aspect of the model is that it can be solved exactly within DMFT---both in equilibrium and in nonequilibrium.
\index{Falicov-Kimball model}

\index{dynamical mean-field theory!impurity solver}
We will not go through the details of the equation of motion for the Green's functions in the Falicov-Kimball model to show how one can solve it. The procedure is most efficiently performed using a path-integral formulation because the time-dependent field on the impurity cannot be easily expressed in terms of a Hamiltonian (unless one introduces a bath for the impurity which does this). Instead, one can employ functional methods to exactly compute the functional derivative of the partition function with respect to the dynamical mean-field, which then is employed to extract the Green's function. Then, because the $f$-electron number is conserved in the model, we obtain the Green's function of the impurity by summing over appropriately weighted combinations of the solution with no $f$-electrons and with one $f$-electron. The end result is that the impurity Green's function for the Falicov-Kimball model in an effective medium is the following:
\begin{equation}
G_0^c(t,t^\prime)=(1-w_1)G_0^c(t,t^\prime)+w_1\left [ \left ( G_0^c\right )^{-1}-U{\mathbb I}\right ]^{-1}_{t,t^\prime}.
\end{equation}
The symbol $w_1=\langle f^\dagger f\rangle$ is the filling of the localized electrons (per lattice site). Since the $f$-electron number operator commutes with ${\mathcal H}$, the localized electron number is a constant of the motion and does not change with time, even in the presence of large electric fields. Note that we will work at half-filling where the electron densities for the conduction electrons and the localized electrons are each equal to $1/2$. This point corresponds to $\mu=U/2$ and $E_f=-U/2$. Details not presented here can be found in the original literature~\cite{freericks1,freericks2}.
\index{dynamical mean-field theory!impurity solver}

\section{Numerical strategies to calculate the Green's function}

Now we need to sort out just how we put this algorithm on a computer and implement it. This discussion closely follows Ref.~\cite{freericks2}. As mentioned above, the challenge is that we are working with continuous matrix operators, with integration over the contour corresponding to matrix multiplication of these operators. Such objects cannot be directly put onto a computer. Instead, we have to first discretize the problem, replacing the continuous operators by 
discrete matrices. There are a few subtle points that are involved corresponding to the integration measure over the contour when we do this, but we  will precisely describe how we achieve this below. In order to recover the results for the continuous matrix operators, we need to extrapolate the discretized results to the limit where the discretization size goes to zero. This is done in the most mundane way possible---simply repeat the calculation for a number of different $\Delta t$ values and use Lagrange interpolation as an extrapolation formula to the $\Delta t\to 0$ limit. We finally check sum rules of the data to verify that the extrapolation procedure worked properly. Details follow.

\index{Kadanoff-Baym-Keldysh contour}
We first need to be concrete about how the discretization is performed. This involves $N_t$ points on the upper real branch (ranging from $t_{min}$ to $t_{max}-\Delta t$), $N_t$ points on the lower real branch (ranging from $t_{max}$ to $t_{min}+\Delta t$), and 100 points along the imaginary axis (ranging from $t_{min}$ to $t_{min}-i\beta+0.1i$, with $\beta=10$); hence $\Delta t=(t_{max}-t_{min})/N_t$. We often find that fixing the number of points on the imaginary time branch rather than scaling them in the same fashion as on the real axis does not cause any significant errors, but it allows the calculations to be performed with fewer total number of points.
The discrete time values on the contour become
\begin{eqnarray}
 t_j&=&-t_{min}+(j-1)\Delta t,\quad\quad\quad 1\le j\le N_t,\label{eq: time_discrete}\\
&=&t_{max}-(j-N_t-1)\Delta t,\quad  N_t+1\le j\le 2N_t,\nonumber\\
&=&t_{min}-0.1i(j-2N_t-1),~ 2N_t+1\le j\le 2N_t+100,
\nonumber
\end{eqnarray}
where we used the fact that the discretization along the imaginary axis is always fixed at $\Delta\tau=0.1$ in our calculations (and we pick the initial temperature to be $T=0.1t^*$ or $\beta=10$).  We use a leftpoint rectangular integration rule for discretizing integrals over the contour, which is implemented as follows:
\begin{equation}
 \int_c dt f(t)=\sum_{i=1}^{2N_t+100}W_if(t_i),
\label{eq: integration_rule}
\end{equation}
where the weights satisfy
\begin{eqnarray}
 W_j&=&\Delta t,\quad\quad\quad 1\le j\le N_t,\nonumber\\
&=&-\Delta t,\quad\quad  N_t+1\le j\le 2N_t,\nonumber\\
&=&-0.1i,\quad\quad 2N_t+1\le j\le 2N_t+100.
\label{eq: integration_weights}
\end{eqnarray}
The leftpoint integration rule evaluates the function at the ``earliest'' point (in the sense of time ordering along the contour, see Fig.~\ref{fig:contour}) in the time interval that has been discretized for the quadrature rule (which is the left hand side of the interval when we are on the upper real branch and right hand side when we are on the lower real time branch).
\index{Kadanoff-Baym-Keldysh contour}

\index{Green's function!boundary condition}
One of the important aspects of many-body Green's functions is that they satisfy a boundary condition. This is what determines whether the Green's function is bosonic or fermionic. For example, you should already be familiar with the fact that the thermal Green's functions are antiperiodic on the imaginary time axis. The contour-ordered Green's function also satisfies a boundary condition where we identify the points $t_{min}$ with $t_{min}-i\beta$. One can show from the definition of the contour-ordered Green's function and the invariance of the trace with respect to the ordering of operators that $G^c_{ii\sigma}(t_{min},t^\prime)=-G^c_{ii\sigma}(t_{min}-i\beta,t^\prime)$
and $G^c_{ii\sigma}(t,t_{min})=-G^c_{ii\sigma}(t,t_{min}-i\beta)$.  The proof is identical to how it is done for the thermal Green's functions. It involves cyclically moving a creation operator from the left to the right and commuting the $e^{-\beta \mathcal H}$ term through it which employs the fact that a cyclic permutation of elements in the product of a trace does not change the value of the trace.
\index{Green's function!boundary condition}

\index{delta function!contour}
The delta function changes sign along the negative real time branch, and
is imaginary along the last branch of the contour in order to satisfy the property that $\int_c dt^\prime \delta_c(t,t^\prime)f(t^\prime)=f(t)$.  In addition, we find that the numerics work better if the definition of the delta function is done via ``point splitting''  (when we calculate the inverse of a Green's function) so that the delta function does not lie on the diagonal, but rather on the first subdiagonal matrix (in the limit as $\Delta t\rightarrow 0$ it becomes a diagonal operator). Because we identify the times $t_{min}$ and $t_{min}-i\beta$, the point splitting approach to the definition of the delta function allows us to incorporate the correct boundary condition into the definition of the discretized delta function. Hence, we define the discretized delta function in terms of the quadrature weights, in the following way
\begin{eqnarray}
 \delta_c(t_i,t_j)&=&\frac{1}{W_i}\delta_{ij+1}, \quad{\rm for~integration~over~}j,\label{eq: delta_cont_defa}\\
&=&\frac{1}{W_{i-1}}\delta_{ij+1},~{\rm for~integration~over~}i,
\label{eq: delta_cont_defb}
\end{eqnarray}
where $t_i$ and $t_j$ are two points on the discretized contour as described in Eq.~(\ref{eq: time_discrete}), and $W_i$ are the quadrature weights described in Eq.~(\ref{eq: integration_weights}). We have a different formula for integration over the first variable versus integration over the second variable because we are using the leftpoint quadrature rule. Note that the formulas in Eqs.~(\ref{eq: delta_cont_defa}) and (\ref{eq: delta_cont_defb}) hold only when $i\ne 1$. When $i=1$, the only nonzero matrix element for the discretized delta function is the  $1,j=2N_t+100$ matrix element, and it has a sign change due to the boundary condition that the Green's function satisfies.
The discretization of the derivative of the delta function on the contour is more complex. It is needed to determine the inverse of the $G_0^c$ for the impurity.  The derivative is calculated by a two-point discretization that involves the diagonal and the first subdiagonal.  Since all we need is the discrete representation of the operator $[i\partial^c_t+\mu]\delta_c(t,t^\prime)$, we summarize the discretization of that operator as
follows
\begin{equation}
 [i\partial_t+\mu]\delta_c(t_j,t_k)=i\frac{1}{W_j}M_{jk}\frac{1}{W_k},
\end{equation}
with the matrix $M_{jk}$ satisfying
\begin{equation}
 M_{jk}=\left (
\begin{smallmatrix}
 1 & 0 & 0 & & & & &...& & & & & 1+i\Delta t \mu\\
-1-i\Delta t \mu & 1 & 0 & & & & & ... & & & & & 0\\
0 & -1-i\Delta t \mu & 1 & 0 \\
 & & & \ddots \\
& & & 0 & -1+i\Delta t\mu & 1 & 0\\
& & & & 0 & -1+i\Delta t\mu & 1\\
& & & & & & &\ddots\\
& & & & & & & & -1-\Delta\tau\mu & 1\\
& & & & & & & & & & \ddots\\
& & & & & & & & & & & -1-\Delta\tau\mu& 1
\end{smallmatrix}
\right );
\end{equation}
here $\Delta\tau=0.1$. The top third of the matrix corresponds to the upper real branch, the middle third to the lower real branch and the bottom third to the imaginary branch.
Note that the operator $[i\partial_t^c+\mu]\delta_c$ is the inverse operator of the Green's function of a spinless electron with a chemical potential $\mu$.  Hence the determinant of this operator must equal the partition function of a spinless electron in a chemical potential $\mu$, namely $1+\exp[\beta\mu]$. This holds because ${\rm det}G^{-1}_{non}={\mathcal Z}_{non}$ for noninteracting electrons. Taking the determinant of the matrix $M_{jk}$ (by evaluating the minors along the top row) gives
\begin{eqnarray}
 \det M&=&1 +(-1)^{2N_t+N_\tau}(1+i\Delta t\mu)(-1-i\Delta t\mu)^{N_t-1}\nonumber\\
&\times&(-1+i\Delta t\mu)^{N_t}(-1-\Delta\tau\mu)^{N_\tau},\nonumber\\
&\approx&1+(1+\Delta\tau\mu)^{N_\tau}+O(\Delta t^2),
\label{eq: det}
\end{eqnarray}
which becomes $1+\exp[\beta\mu]$ in the limit where $\Delta t,\Delta\tau\rightarrow 0$ ($N_\tau$ is the number of discretization points on the imaginary axis).  This shows the importance of the upper right hand matrix element of the operator, which is required to produce the correct boundary condition for the Green's function. It also provides a check on our algebra. In fact, we chose to point-split the delta function when we defined it's discretized matrix operator precisely to ensure that this identity holds.
\index{delta function!contour}

We also have to show how to discretize the continuous matrix operator multiplication and how to find the discretized approximation to the continuous matrix operator inverse.  As we mentioned above, the continuous matrix operator is described by a number for each entry $i$ and $j$ in the discretized contour. We have to recall that matrix multiplication corresponds to an integral over the contour, so this operation is discretized with the integration weights $W_k$ as follows
\begin{equation}
 \int_c d\bar t A(t,\bar t)B(\bar t,t^\prime)=\sum_k A(t_i,t_k)W_kB(t_k,t_j).
\label{eq; matrix_multiply}
\end{equation}
Thus we must multiply the columns (or the rows) of the discrete matrix by the corresponding quadrature weight factors when we define the discretized matrix.  This can be done either to the matrix on the left (columns) or to the matrix on the right (rows). To calculate the inverse, we recall the definition of the inverse for the continuous matrix operator
\begin{equation}
 \int_c d\bar t A(t,\bar t)A^{-1}(\bar t, t^\prime)=\delta_c(t,t^\prime),
\label{eq: matrix_inverse}
\end{equation}
which becomes the following
\begin{equation}
\sum_k A(t_i,t_k)W_kA^{-1}(t_k,t_j)=\frac{1}{W_i}\delta_{ij},
\label{eq: matrix_inverse_discrete}
\end{equation}
in its discretized form. Note that
we do not need to point-split the delta function here.  Hence, the inverse of the matrix is found by inverting the matrix $W_iA(t_i,t_j)W_j$, or, in other words, we must multiply the rows and the columns by the quadrature weights before using conventional linear algebra inversion routines to find the discretized version of the continuous matrix operator inverse. This concludes the technical details for how to discretize and work with continuous matrix operators.

\index{Green's function!local}
The next technical aspect we discuss is how to handle the calculation of the local Green's function. Since the local Green's function is found from the following equation:
\begin{equation}
G_{ii}^c(t,t^\prime)=\frac{1}{\pi t^{*2}}\int d\epsilon \int d\bar\epsilon\,e^{-\frac{\epsilon^2}{t^{*2}}-\frac{\bar\epsilon^2}{t^{*2}}}\left [\left ({\mathbb I}-G_{\epsilon,\bar\epsilon}^{c,non}\Sigma^c\right )^{-1}G_{\epsilon,\bar\epsilon}^{c,non}\right ]_{t,t^\prime}
\end{equation}
with the noninteracting Green's function discussed earlier and given by
\begin{equation}
G_{\epsilon,\bar\epsilon}^{c,non}(t,t^\prime)=i\left[ f(\epsilon-\mu)-\theta_c(t,t^\prime)\right ]
e^{-i\int_{t^\prime}^t d\bar t \left [ \cos\left ( A(\bar t)\right )\epsilon-\sin\left (A(\bar t)\right )\bar\epsilon\right ]}.
\end{equation}
Note that we must have $A(t)=0$ before the field is turned on. Of course an optical pump pulse also requires $\int_{-\infty}^\infty A(t)dt=0$, because a traveling electromagnetic wave has no dc field component. We did not compute the matrix inverse of the noninteracting Green's function because re-expressing the formula in the above fashion makes the computation more efficient (because matrix multiplication requires less operations than a matrix inverse does).
\index{Green's function!local}

This step is the most computationally demanding step because we are evaluating the double integral of a matrix-valued function and we need to compute one matrix inverse for each integrand. Fortunately, the computation for each $(\epsilon,\bar\epsilon)$ pair is independent of any other pair, so we can do this in parallel with a master-slave algorithm. In the master-slave algorithm, one CPU controls the computation, sending a different $(\epsilon,\bar\epsilon)$ pair to each slave to compute its contribution to the integral and then accumulating all results. The sending of the pairs to each slave is simple. One has to carefully manage the sending of the results back to the master, because the data is a large general complex matrix and they are all being received by one node. This is precisely the situation where a communication bottleneck can occur.

\index{Gaussian integration}
In order to use as few integration points as possible, we employ Gaussian integration, because the bare density of states for both the $\epsilon$ and the $\bar\epsilon$ integrals is a Gaussian. We found it more efficient to average the results with two consecutive numbers of points in the Gaussian integration (like 100 and 101) instead of just using 200 points (which would entail twice as much computation). 
\index{Gaussian integration}

The rest of the DMFT loop requires serial operations and is performed on the master to reduce communications. One does have to pay attention to convergence. In metallic phases, and with strong fields, the code will converge quickly But when the field is small or the interactions large, one might not be able to achieve complete convergence. Often the data is still good, nevertheless. One also should not make the convergence criterion too stringent. Usually 4 digits of accuracy is more than enough for these problems. In most cases, one will need to iterate as many as 50-100 times for hard cases. But many results can be obtained  with ten iterations or less. 

Finally, one has to repeat the calculations for different discretizations and extrapolate to the $\Delta t\to 0$ limit. In order to use as much data as possible, it is best to first use a shape-preserving Akima spline  to put all data on the same time grid and then use Lagrange interpolation as an extrapolation method to the $\Delta t\to 0$ limit. It is critical that one employs a shape-preserving spline, otherwise it will ruin your results. In addition, we  find quadratic extrapolation usually works best, but sometimes had to look at higher-order extrapolations. It is best to extrapolate the data after the desired quantity has been determined as a function of $t$ for all times on the grid. For example, one would first determine the current by using the lesser Green's function and summing over all momentum for each time and then extrapolate the final current values as a function of time instead of extrapolating the Green's functions, since doing the latter requires a two-dimensional Lagrange extrapolation formula and uses large amounts of memory.

\begin{figure}[t!]
\centering
\includegraphics[width=4.5in]{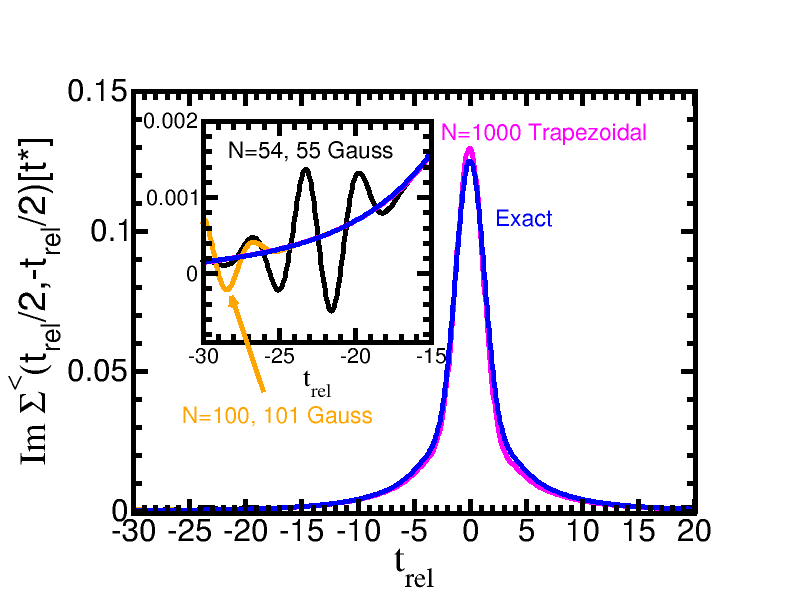}
\caption{Local lesser self-energy for the Falicov-Kimball model at half-filling and $\beta=1$ and $U=1$. The timestep was $\Delta t=0.05$ and the time range ran from $-15\le t_{rel}\le 15$. The blue curve is the exact result and the other colors are different methods for integration over the noninteracting density of states. The inset shows a blow up of the region at large negative times, where the different integration methods start to break down. This figure is adapted from Ref.~\cite{freericks_ugc}.}
\label{fig:eq_se}
\end{figure}

\index{sum rules}
Finally we mention that the accuracy can be tested by calculating spectral moment sum rules. These are known exactly for the retarded Green's function (through third-order).  At half-filling they simplify tremendously. We will explore how this works when we discuss some examples in the next section. It is important to perform this test on {\it all} of your data. It is easy to do and ensures the data is high quality.
\index{sum rules}

We end this section with a discussion of other methods. Tran showed how one can reduce the size of the matrices in half by working with the retarded and lesser Green's functions only~\cite{tran}, but such an algorithm is harder to code and you need to solve for twice as many Green's functions, so it reduces the computational time by only a factor of two (it may allow for slightly larger matrices to be handled as well). There is an alternative way to discretize and solve this problem that integrates the Dyson equation in time in a causal fashion, producing the Green's function along the boundary of the matrix as time increases~\cite{contour_alt}. It is very efficient, but requires you to be able to determine the self-energy in a causal fashion from your data at all {\it earlier} times. This is always possible when perturbation theory is used to determine the self-energy. But, while there should be a way to do this for the Falicov-Kimball model, the explicit method has not yet been discovered.  For the Hubbard model, we have no exact methods available.

\section{Examples}

We now illustrate how all of this machinery works. The idea is not to provide an exhaustive review of the research work that has been completed, but rather to illustrate how these ideas are concretely put into action. As such, we will pick and choose the examples more for their didactic features than for their importance, or perhaps even their physical relevance.

\index{self-energy}
\begin{figure}[htb]
\centering
\includegraphics[width=4.5in]{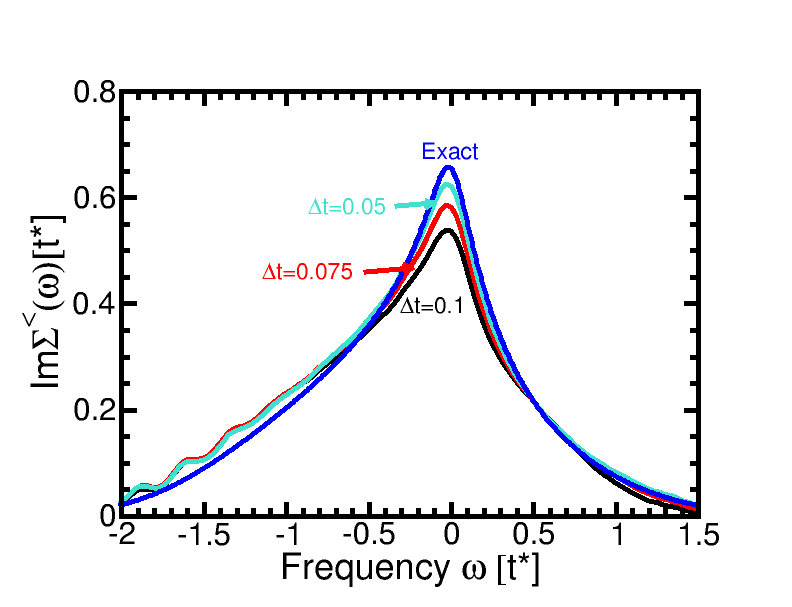}
\caption{Local lesser self-energy for the Falicov-Kimball model at half-filling and $\beta=1$ and $U=1$. Here, we Fourier transform the data from time to frequency. We use different time discretizations for the different curves, while the density of states integration used $N=54$ and $N=55$ points for Gaussian integration. The oscillations on the discrete calculations come from the oscillations in time shown in Fig.~\ref{fig:eq_se}. This figure is adapted from Ref.~\cite{freericks_ugc}.}
\label{fig:eq_se_w}
\end{figure}
\index{self-energy}

\index{self-energy}
To start, we focus on equilibrium. The DMFT for the Falicov-Kimball model was solved exactly in equilibrium by Brandt and Mielsch~\cite{brandt_mielsch1,brandt_mielsch2} and is summarized in a review~\cite{freericks_review}. We take the exact result as a function of frequency and Fourier transform to time. Then we compare that result with the result that comes out of the nonequilibrium formalism, to understand its accuracy. In all of this work, we solve the Falicov-Kimball model on a hypercubic lattice at half filling. In Fig.~\ref{fig:eq_se}, we plot the (local) lesser self-energy $\Sigma^<(t)$. The calculation used a fixed $\Delta t=0.05$ and has $U=0.5$. The different curves are for various different integration schemes over the noninteracting density of states. In one case, we average the $N=54$ and $N=55$ Gaussian formulas. In another, we do the same, but with $N=100$ and $N=101$. We also compare to a much denser trapezoidal formula with 1000 points. Additionally, we plot the exact result. One can see the different integration schemes are quite close to each other for times near zero, but they begin to deviate at large times. The inset focuses on large negative time and one can quickly conclude that there is a maximum absolute value of time for which the results are accurate for any integration scheme. Beyond that, the system starts to generate increasing amplitude oscillations. The disagreement at short times from the exact result stems from the fact that these calculations are at a fixed $\Delta t$---no scaling to the $\Delta t\to 0$ limit were taken. This gives a sense of the accuracies we can hope to attain with this approach.
\index{self-energy}

\index{Green's function}
\begin{figure}[htb]
\centering
\includegraphics[width=4.5in]{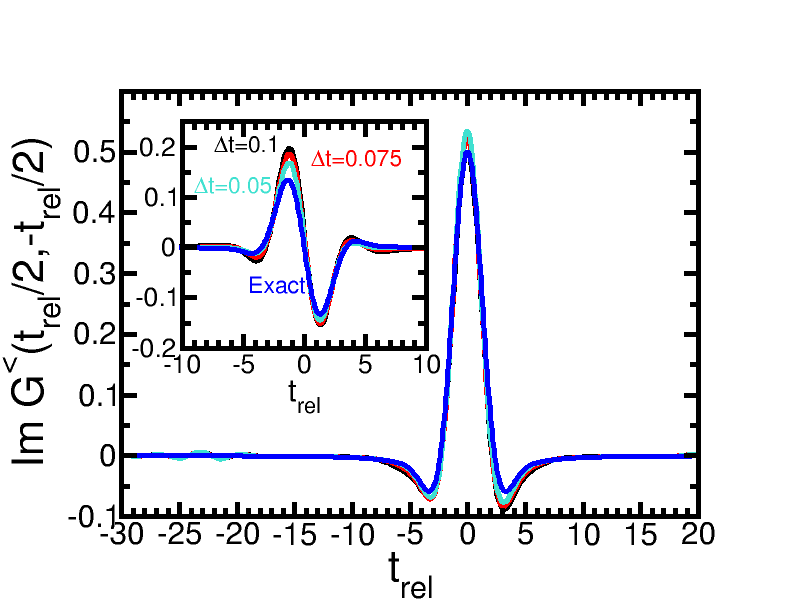}
\caption{Local lesser Green's function for the Falicov-Kimball model at half-filling and $\beta=1$ and $U=1$.The parameters are the same as the previous figure. The main plot is the imaginary part (which becomes symmetric for the exact result, and the inset is the real part, which becomes antisymmetric.  This figure is adapted from Ref.~\cite{freericks_ugc}.}
\label{fig:eq_gf}
\end{figure}
\index{Green's function}

\index{self-energy}
In Fig.~\ref{fig:eq_se_w}, we plot the Fourier transform of the time data as a function of frequency. The number of Gaussian points used is $N=54$, 55. The time steps are varies and one can see that they are approaching the exact result. If we use Lagrange extrapolation here, we would get quite close to the exact result, but we do not include that plot here because it would be too many lines close to each other. The oscillations in the tail of the data in Fig.~\ref{fig:eq_se} is responsible for the oscillations at negative frequency seen in the data. Those oscillations will remain even after scaling to the $\Delta \to 0$ limit. One can clearly see how the extrapolation method works for this case.
\index{self-energy}

\index{Green's function}
In Fig.~\ref{fig:eq_gf}, we do a similar plot in the time domain for the lesser Green's function. Note that the real part of the lesser Green's function (inset) is an odd function when we are at half-filling due to particle-hole symmetry. Similarly, the imaginary part in the main panel is even for the exact result. One can clearly see how the extrapolation will work to approach the exact result if we did the full extrapolation of this data. 
\index{Green's function}

\index{density of states!equilibrium}
\begin{figure}[htb]
\centering
\includegraphics[width=4.5in]{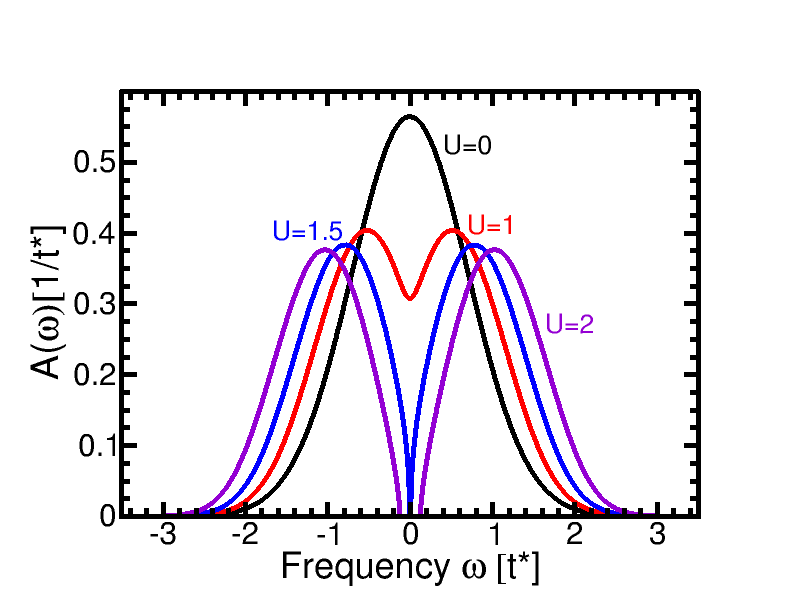}
\caption{Local density of states for the Falicov-Kimball model in equilibrium and at half-filling. The calculations are in equilibrium, where the density of states is temperature independent in the normal state. The curves are for different $U$. One can see as $U$ increases we cross through a Mott transition at $U\approx \sqrt{2}$.}
\label{fig:eq_dos}
\end{figure}
\index{density of states!equilibrium}

\index{Mott transition}
Now that we have gotten our feet wet with the numerics, we are ready to discuss some physics. At half-filling, there are enough conduction electrons to fill half of the sites of the lattice and similarly for the localized electrons. In this case, if the repulsion between the two species is large enough, they will avoid each other and the net effect is that the system becomes an insulator because charge motion is frozen due to the high energy cost for double occupancy. We can see this transition occur in Fig.~\ref{fig:eq_dos}, which is the local density of states at half-filling for different $U$ values. As $U$ is increased we evolve from the initial result, which is Gaussian for no interactions to results where a ``hole'' is dug into the density of states until if opens a gap at a critical value of $U$ called the Mott transition. Beyond this point, the system has a gap to charge excitations.
\index{Mott transition}

\index{density of states!nonequilibrium}
\begin{figure}[t!]
\centering
\includegraphics[width=3.0in]{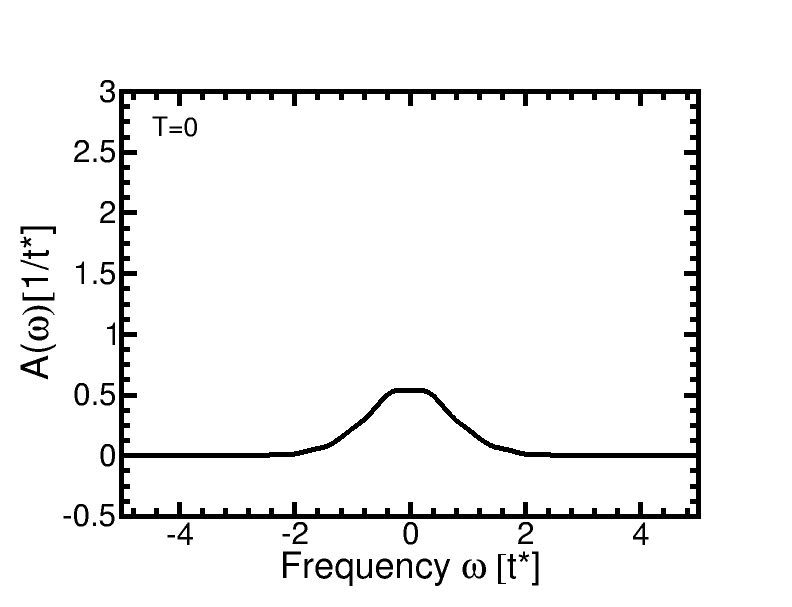}\includegraphics[width=3.0in]{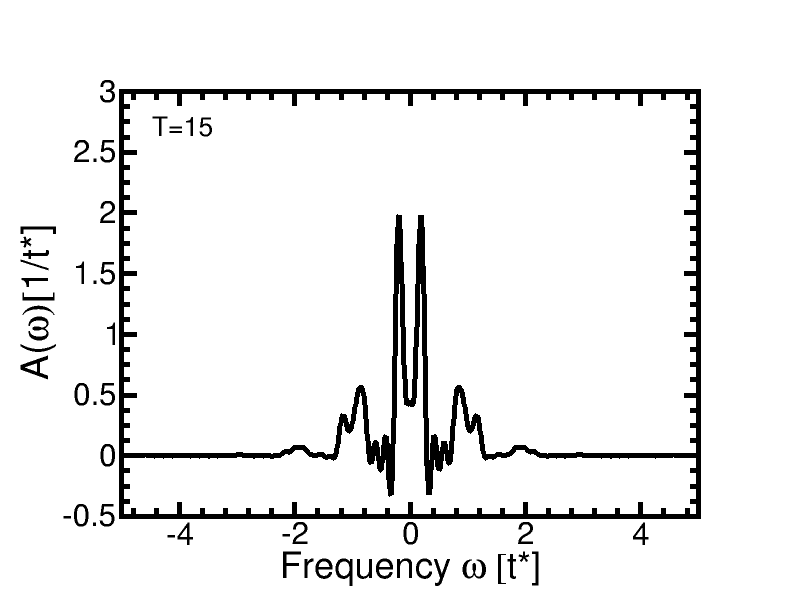}
\includegraphics[width=3.0in]{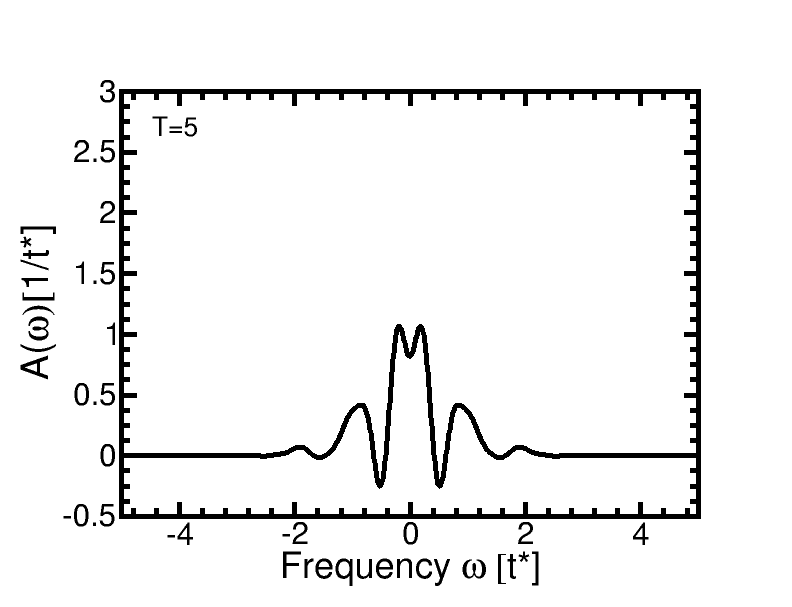}\includegraphics[width=3.0in]{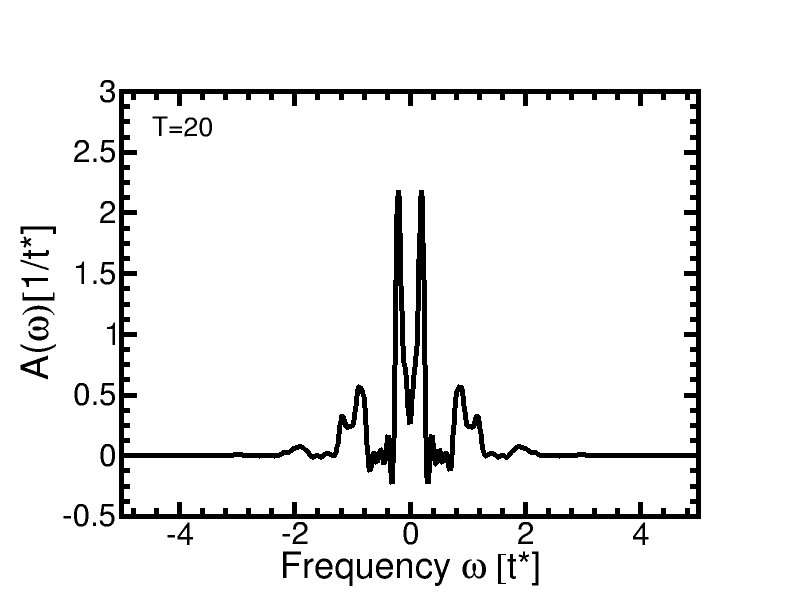}
\includegraphics[width=3.0in]{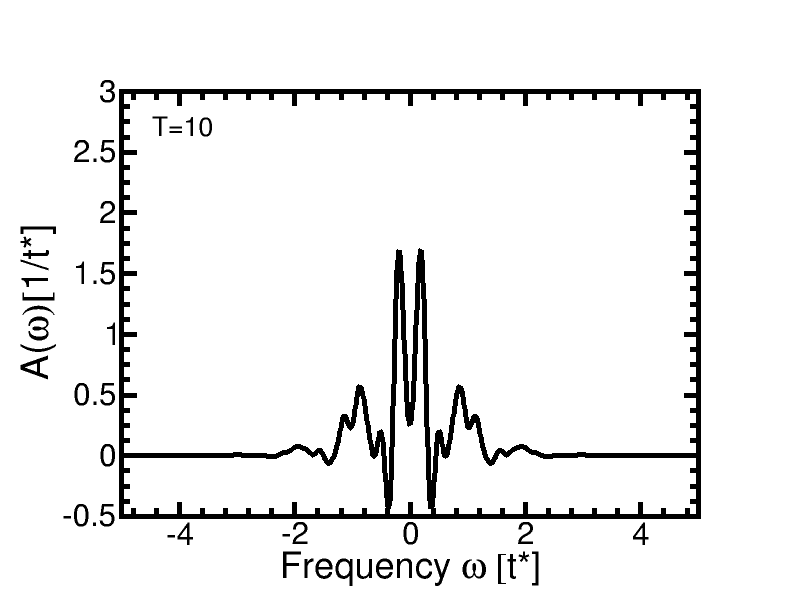}\includegraphics[width=3.0in]{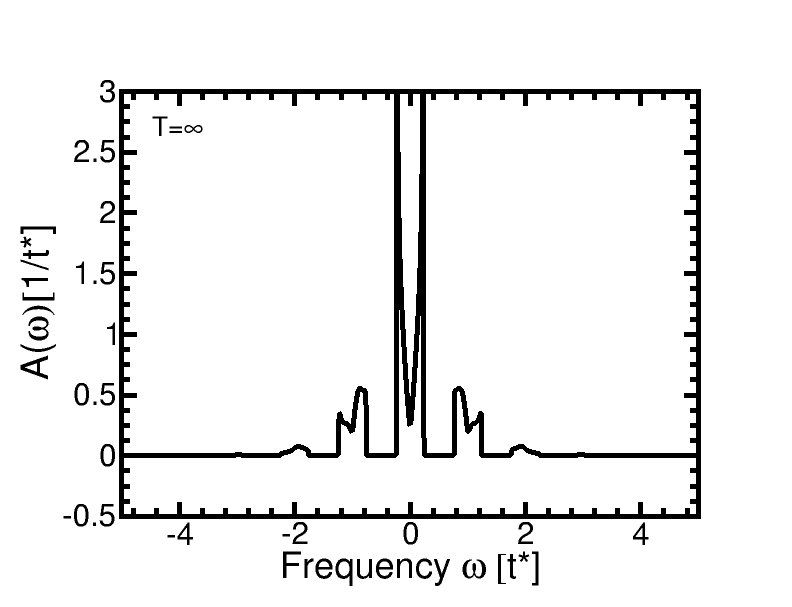}
\caption{Transient local density of states for the Falicov-Kimball model in a dc electric field with $U=0.5$, $\beta=0.1$ and $E=1$ at half-filling. The different panels are labelled by the average time of the spectra. One can see that quite quickly after the dc field is turned on, the retarded Green's function approaches the steady state. These results are adapted from~\cite{freericks_turk_rev}}
\label{fig:neq_dos}
\end{figure}
\index{density of states!nonequilibrium}

\index{density of states!nonequilibrium}
Next, we illustrate how the density of states evolves in a transient fashion in time after a dc field is turned on. In Fig.~\ref{fig:neq_dos}, we show a series of frames at different time snapshots that plot the transient density of states (imaginary part of the local retarded Green's function) as a function of frequency. One can see that the system starts off in a near Gaussian density of states and then develops features that are quite complex. The infinite-time limit is solved using Floquet nonequilibrium DMFT theory~\cite{freericks_neq_hubbard,freericks_joura,aoki_floquet}; unfortunately, we are not able to describe the details for how that problem is solved. But there are a few important features to notice. Both before the field is applied and long after it has been applied, the system has a positive semidefinite density of states. This is the same as in equilibrium and it allows the density of states to be interpreted probabilistically. But in the transient regime, it often becomes negative. Furthermore, because the density of states is determined via a Fourier transformation, one cannot just say it corresponds to a spectra at a specific average time. Instead it senses times near the average time governed by how rapidly the Green's function decays in the time domain.
\index{density of states!nonequilibrium}

\index{density of states!nonequilibrium}
\index{Wannier-Stark ladder}
There is a lot of physics in these figures. If there was no interactions, the system would undergo Bloch oscillations, because the electrons are not scattering. The density of states then is the Fourier transform of a periodic function, which leads to delta function peaks forming the so-called Wannier-Stark ladder, with the separations given by the dc field $E$ that is applied to the system. When interactions are added in, the Wannier-Stark ladder is broadened, but also split by the interaction $U$. This occurs because a delta function has a zero bandwidth and hence is highly susceptible to the Mott transition, even for relatively small interactions.
\index{density of states!nonequilibrium}
\index{Wannier-Stark ladder}

\index{current}
\index{Bloch oscillations}
\begin{figure}[htb]
\centering
\includegraphics[width=4.5in]{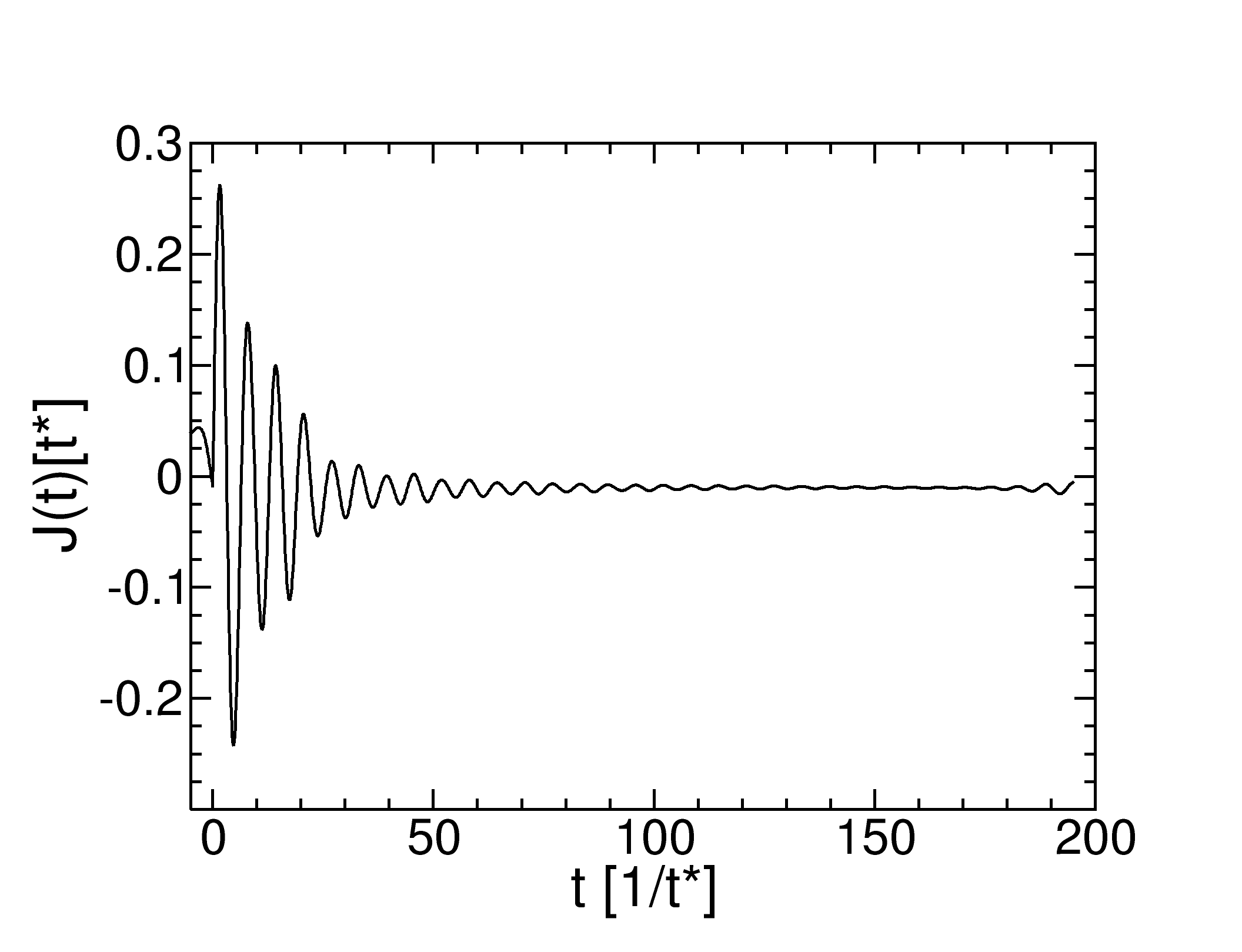}
\caption{Transient current for $E=1$, $U=0.5$ and $\beta=0.1$. These results use $\Delta t=0.1$ and are not scaled. Notice how they start off as a Bloch oscillation, but are damped by the scattering due to the interaction $U$. They die off and then start to have a recurrence at the longest times simulated. At the earliest times the current is nonzero simply because we have not scaled the data. Scaling is needed to achieve a vanishing current before the field is turned on.  These results are adapted from Ref.~\cite{freericks2}.}
\label{fig:current}
\end{figure}
\index{current}
\index{Bloch oscillations}

\index{current}
\index{Bloch oscillations}
Next, we move on to examining the transient current. We use the same  case we have been examining throughout this brief summary---$U=0.5$ and $\beta=0.1$. Here the dc field is $E=1$. One can see the current starts of as a weakly damped Bloch oscillations (underdamped). It dies off and remains quiescent for some time and then starts to recur at the longest times. The characteristic Bloch oscillations occurs for both metals and insulators, but it is damped much more rapidly in insulators because they interact more strongly. This is one of the common observables measured in a nonequilibrium experiment. But it is not measured directly, because oscilloscopes are not fast enough to see them.
\index{current}
\index{Bloch oscillations}

\index{sum rules}
\begin{figure}[htb]
\centering
\includegraphics[width=3.3in]{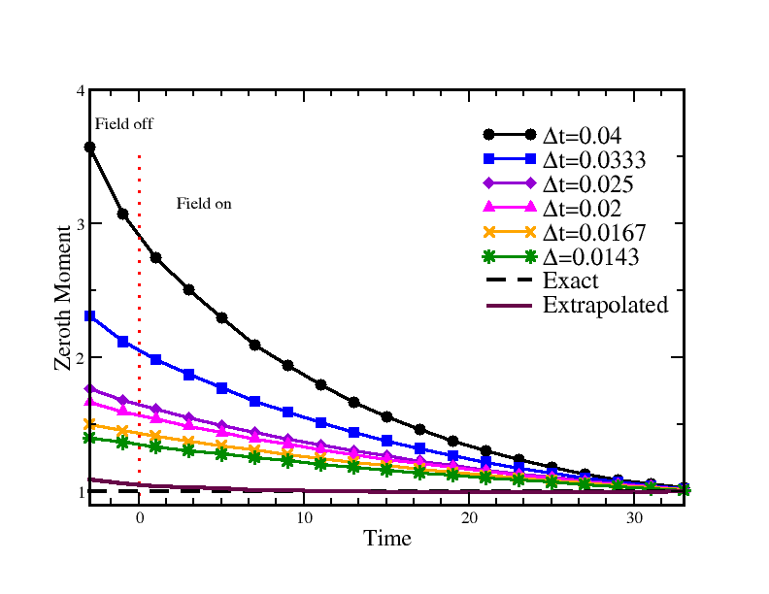}\includegraphics[width=3.3in]{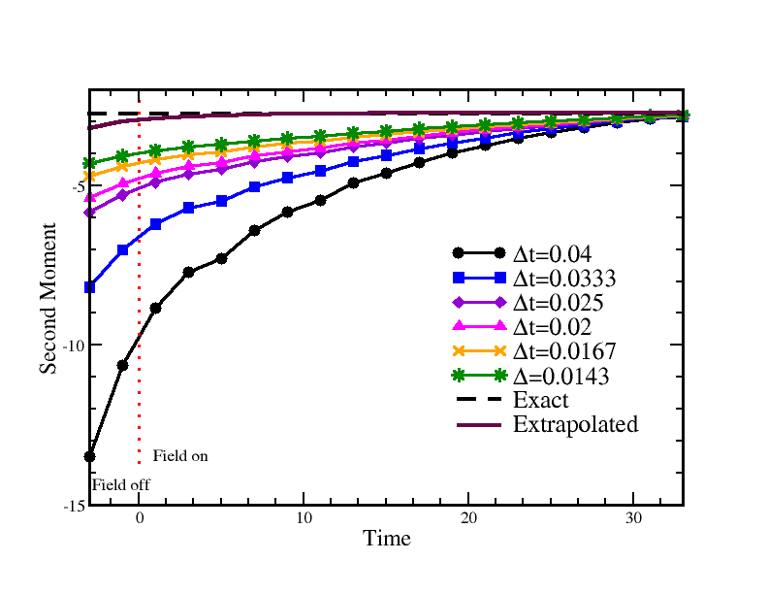}
\caption{Sum rules for the local retarded Green's function for $E=1$ and $U=2$. Here, we illustrate how one can use sum rules to verify the scaling to $\Delta t\to 0$ has been done accurately. The zeroth moment sum rule equals 1 and the second moment sum rule equals $-(1/2+U^2/4)$. We plot the raw data versus time, the exact result, and the extrapolated result. One can see that even if the raw data was off by a huge amount, the final extrapolated data works extremely well.  These results are adapted from Ref.~\cite{ugc2}.}
\label{fig:sum_rules}
\end{figure}
\index{sum rules}

\index{sum rules}
Finally, we show how the sum rules hold and illustrate why it is important to scale results to the $\Delta t\to 0$ limit. We pick a case which is challenging to calculate $E=1$ and $U=2$. This is in the Mott insulating phase, where the numerics are much more difficult to manage. We primarily use the sum rules to indicate whether the calculations are accurate enough that they can be trusted. As shown in Fig.~\ref{fig:sum_rules}, we can see that the raw data can be quite bad, but the final scaled result ends up being accurate to 5\% or less! Note how the results are worst on the equilibrium side (to the left) than on the nonequilibrium side (to the right) and they approach the exact results on the nonequilibrium side. Hence, these calculations are most accurate in moderate to large fields.
\index{sum rules}

This ends our short representative tour of some numerical results that can be calculated with this nonequilibrium DMFT approach.

\section{Conclusions}

These lecture notes have been somewhat brief due to the page constraints of this contribution (and the time constraints of its author). Nevertheless, I have tried to present enough information here that you can follow the logic, reasoning, and even develop the formalism for yourself if you want to engage in these types of calculations. The field of nonequilibrium many-body physics is wide open. There are new and interesting experiments performed every day and the theory to describe them still needs further development. We do have a number of successes. The approach can be used for optical conductivity~\cite{noneq_rmp_rev,lex_oc}, time-resolved angle-resolved photoemission~\cite{freericks_trarpes}, Raman scattering~\cite{freericks_raman}, x-ray photoemission spectroscopy and x-ray absorption spectroscopy~\cite{freericks_xas}, and resonant inelastic x-ray scattering~\cite{freericks_rixs}. The challenge is always about how far out in time can a simulation go. 
If you have  good idea for a nonequilibrium solver for a Hubbard impurity, I encourage you to give it a try. We really need it. I also hope you will enjoy working with many-body Green's functions in the future. They are truly wonderful!

\section*{Acknowledgments}
This work was supported by the Department of Energy, Office of Basic Energy Sciences, Division of Materials Science and Engineering under contract number DE-FG02-08ER46542. It was also supported by the McDevitt bequest at Georgetown University. I would also like to thank the organizers for the invitation to contribute to this school.


\clearchapter


\end{document}